\documentstyle[prb,aps,epsf]{revtex}

\begin{document}
\draft

\title{\bf Reptation of star polymers in a network :\\
Monte Carlo results of diffusion coefficients}

\author{G.T. Barkema}
\address{Theoretical Physics, Utrecht University, Princetonplein 5,
3584 CC Utrecht, The Netherlands}
\author{A. Baumgaertner}
\address{Forum Modellierung, Forschungszentrum, 52425 J\"ulich, Germany}

\date{received 25 August 1998; to appear in {\it Macromolecules}}

\maketitle
\begin{abstract}
We report on Monte Carlo results of diffusion coefficients of lattice
star polymers trapped inside a fixed network (de Gennes model).  It is
found that our data are in agreement with the Helfand-Pearson
exponential factor $\alpha = 0.29$.
For the pre-exponential power law exponent we find $\beta=2$.
In contrast to
existing theoretical predictions, we find that the number of arms $f$
leads to a pre-exponential factor of the form $\exp(-0.75 f)$.
\end{abstract}


\vskip2pc


\section{Introduction}

Entangled star polymers have the unique property that the viscosity
does not increase with a power law of the molecular weight, as found
for linear polymers, but instead increases exponentially.  This
experimental result was first reported by Kraus and Gruver \cite{Kraus}
and subsequently confirmed by many others \cite{Fetters}.  Theoretical
explanations have been given based on extensions of the reptation model
\cite{deGennes,Doi,Graessley,Needs,Pearson}.  An essential assumption
made by adapting the reptation model to entangled stars is that the
dominating mechanism for translational motion is the retraction of ends
of the arms along their average contour, with the simultaneous
projection of unentangled loops into the surrounding matrix.  As the
arms retract the free energy increases due to the loss of
configurational entropy. De Gennes \cite{deGennes} was the first to
explore this problem using a lattice model defined in the next
section.  He calculated the number of walks for each tube length and
concluded that the time required to retract one arm completely would
increase exponentially with the arm length $N$. In a subsequent
approach Doi and Kuzuu \cite{Doi} obtained the time dependence of the
star relaxation. The Doi-Kuzuu theory predicts that the longest
relaxation time $\tau_m$ increases exponentially with the molecular
weight and has a power law prefactor
\begin{equation}
\label{tau}
\tau_m \propto N^{\beta}\;{\rm e}^{-\alpha N}.
\end{equation}
The power law exponent $\beta$ has been predicted by Doi and Kuzuu
\cite{Doi} to $\beta=3$, by Pearson and Helfand \cite{Pearson} to
$\beta=3/2$, and has been estimated by Needs and Edwards \cite{Needs}
using Monte Carlo methods to $\beta \approx 1.9$.  The exponential
factor $\alpha$ has been calculated by Helfand and Pearson~\cite{Helfand}
\begin{equation}
\label{alpha}
\alpha_{HP} = \frac{1}{2} \ln (\frac{q^2}{4 (q-1)}),
\end{equation}
where $q$ is the lattice coordination number. In the present case of
the simple cubic lattice one has $q=6$ and thus $\alpha_{HP} =
0.29389$.  The numerical estimate of $\alpha=0.195$ by Needs and
Edwards~\cite{Needs} is significantly smaller than this.
Since during the time $\tau_m$ the star polymer diffuses by a tube
diameter, which is in the order of unity in our lattice model, one can use
the relation $\tau_m D \sim 1$
between the terminal relaxation time
$\tau_m$ and the diffusion coefficient $D$ in order to obtain
\begin{equation}
\label{diff}
D = c\;N^{-\beta}\;{\rm e}^{-\alpha N}.
\end{equation}
Using Monte Carlo simulations Needs and Edwards \cite{Needs} found
$\beta=0.59$, $\alpha=0.37$ and $c=0.033$ for $f=3$.

With respect to the conflicting estimates of $\beta$ and $\alpha$ we
have attempted to clarify this situation by simulating again the
lattice model proposed by de Gennes.  The advance in computers combined
with a highly optimized multispin coding technique allows us to
simulate over time scales that are several orders of magnitude longer
than previous simulations \cite{Needs}.  Moreover, we have addressed
the dependence of the diffusion coefficient on the number of arms $f$.

\vskip 2cm
\section{model and simulation techniques}

We simulate a model for a polymer in a gel that was introduced by de
Gennes.  In this model \cite{Evans,Needs}, a star polymer with $f$ arms
and arm length $N$ is represented by $fN+1$ monomers on {{a 3D cubic}}
lattice, connected by a sequence of $fN$ steps on lattice edges. One
elementary move in this model is made by randomly selecting one
monomer, and attempting to move it. If all steps connecting the monomer
are located on the same edge, it will randomly move to one of the six
possible lattice sites (the site it is already in, and five new ones).
The rate of any monomer to move to any allowed lattice site is once per
time unit.  A configuration of a three-legged star polymer in the
two-dimensional version of this model is illustrated in Fig.~1. In this
configuration, monomers 1, 2, 3, 6, 7, 9, and 14 can move to three
other locations, the other monomers are frozen.  Note that all our
simulations were done in the three-dimensional model.

\begin{figure}
\epsfxsize=8cm
\begin{center}
\epsfbox{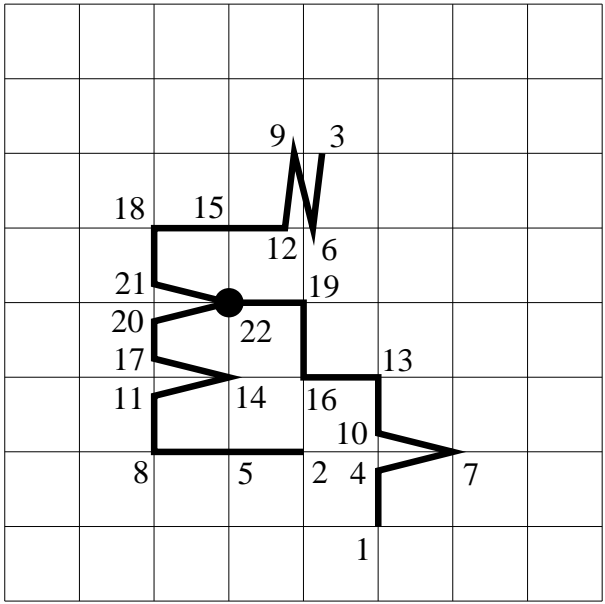}
\caption{Two-dimensional version of the lattice model for a star polymer
with three arms (f=3) and N=7 steps.}
\label{starmodel}
\end{center}
\end{figure}

A straightforward way to store the configuration of the polymer would
be to store the coordinates of all the monomers. However, a more
efficient implementation of the algorithm can be obtained by storing
the coordinates of the central monomer only, plus for each arm the
sequence of steps taken from this center.  In one elementary move in a
direct implementation of the dynamics, randomly one of the $fN+1$
monomers is chosen and proposed to move.

\begin{itemize}
\item{(a)} The central monomer can only move if the first steps of
all arms are in the same direction; in this case, these steps will be
replaced by a randomly chosen direction, and the coordinates of the central
monomer are updated.

\item{(b)} One of the end monomers can always move, and the last step is
replaced by a random direction.

\item{(c)} Any other monomer can only move if the two steps connected
to it are opposites. In this case, these steps are replaced by
a pair of randomly chosen opposite steps.

\end{itemize}

To obtain that each of the $6fN+6$ possible moves is tried with unit rate,
one elementary move corresponds to a time increment of $\Delta t=1/(6fN+6)$.

It turns out that the diffusion coefficient decays exponentially with
arm length, requiring a very efficient implementation of the algorithm
to extract reliable numerical results. We achieve that by using
multispin coding, enabling us to make about seven million elementary
moves per second per processor on a SG 200 workstation. In the past, we
simulated the repton model proposed by Rubinstein \cite{rubinstein}
with the same approach in an electric field~\cite{repton}, and in the
absence of an electric field~\cite{Drepton1,Drepton2}, and de Gennes model for
linear polymers~\cite{krenzlin}. The processors are 64-bit, allowing
for 64 concurrent simulations. For each of the 64 simulations, we store
the coordinates of central monomer as $(x_c, y_c, z_c)^{(0)} \dots
(x_c, y_c, z_c)^{(63)}$.  Suppose that we denote the $k^{th}$ bit of
long integer $w$ as $w^k$, then step $i$ in arm $j$ of simulation $k$
is stored in the three bits $\{a[if+j]^k, b[if+j]^k, c[if+j]^k\}$; if
the step is in the positive $x$-, $y$-, or $z$-direction, this triplet
is equal to \{1,0,0\}, \{0,1,0\} or \{0,0,1\}, respectively; the
negative directions are represented by \{0,1,1\}, \{1,0,1\}, and
\{1,1,0\}, respectively. Note that opposite steps are each other's
binary complement.  The core of the program, that proposes moves of
monomers other than the central and end monomers simultaneously in all
64 simulations, can then be written (in the programming language C) as:

\vskip 0.2in
\noindent
{\tt flip=(a[p]$\wedge$a[q])\&(b[p]$\wedge$b[q])\&(c[p]$\wedge$c[q]);\\
  nf= $\sim$flip;\\
  a[p]=(a[p]\&nf) | (  rnda \&flip);\\
  b[p]=(b[p]\&nf) | (  rndb \&flip);\\
  c[p]=(c[p]\&nf) | (  rndc \&flip);\\
  a[q]=(a[q]\&nf) | (($\sim$rnda)\&flip);\\
  b[q]=(b[q]\&nf) | (($\sim$rndb)\&flip);\\
  c[q]=(c[q]\&nf) | (($\sim$rndc)\&flip);
}
\vskip 0.2in

\noindent
Here, $p$ and $q=p+f$ are steps connected to the same monomer, the
symbols $\wedge$, $\&$, $\sim$ and $|$ denote respectively the
exclusive-OR, AND, NOT and OR operations, and the triplet
$(rnda,rndb,rndc)$ represents a random direction.  Similar statements
can be written down for the central and end monomers.

\section{The diffusion constant}

The star polymers were simulated over a long time $t_{tot}$ (around
$3 \cdot 10^8$ for the longest arm lengths), and the coordinates of the
central monomers were written to a file at regular times. Afterwards, a
quick estimate for the diffusion coefficient from
\begin{equation}
D=\frac{\langle \left( \vec{r}(t_1)-\vec{r}(t_0) \right)^2 \rangle}{6(t_1-t_0)}
\end{equation}
with $t_1=t_{tot}$ and $t_0=0$; with this quick estimate we determined
the time interval $\Delta t$ after which the mean square displacement
of the center equals the arm length $N$, and performed a better
estimate of $D$ using the same equation with $t_1=t_0+\Delta t$,
averaged over all $t_0$. The results are presented in table \ref{Dtable}.

\begin{table}[t]
\caption{The diffusion coefficients $D_f$
of star polymers with $f$ arms and arm lengths $N$.}
\begin{center}
\begin{tabular}{lrrcc}
$N$ & $D_3$ & $D_4$ & $D_5$ & $D_6$ \\
\hline
\multicolumn{2} {c} {}\\
2  & $1.09 (3) \cdot 10^{-2}$ & $2.37 (2) \cdot 10^{-3}$ & $4.70 (2) \cdot
10^{-4}$ & $9.27 (2) \cdot 10^{-5}$ \\
3  & $5.09 (3) \cdot 10^{-3}$ & $1.26 (2) \cdot 10^{-3}$ & $2.90 (3) \cdot
10^{-4}$ & $5.69 (2) \cdot 10^{-5}$ \\
4  & $2.98 (4) \cdot 10^{-3}$ & $8.14 (2) \cdot 10^{-4}$ & $2.02 (4) \cdot
10^{-4}$ & $4.46 (4) \cdot 10^{-5}$ \\
5  & $1.78 (2) \cdot 10^{-3}$ & $5.42 (8) \cdot 10^{-4}$ & $1.50 (2) \cdot
10^{-4}$ & $3.29 (3) \cdot 10^{-5}$ \\
6  & $1.11 (1) \cdot 10^{-3}$ & $3.48 (7) \cdot 10^{-4}$ & $1.08 (1) \cdot
10^{-4}$ & $2.88 (5) \cdot 10^{-5}$ \\
7  & $7.00 (7) \cdot 10^{-4}$ & $2.35 (4) \cdot 10^{-4}$ & $7.82 (5) \cdot
10^{-5}$ & $2.27 (5) \cdot 10^{-5}$ \\
8  & $4.47 (5) \cdot 10^{-4}$ & $1.62 (3) \cdot 10^{-4}$ & $5.47 (4) \cdot
10^{-5}$ & $1.72 (4) \cdot 10^{-5}$ \\
9  & $2.87 (3) \cdot 10^{-4}$ & $1.08 (2) \cdot 10^{-4}$ & $3.93 (6) \cdot
10^{-5}$ & $1.31 (6) \cdot 10^{-5}$ \\
10 & $1.85 (3) \cdot 10^{-4}$ & $7.39 (1) \cdot 10^{-5}$ & $2.89 (7) \cdot
10^{-5}$ & $1.09 (7) \cdot 10^{-5}$ \\
12 & $8.17 (9) \cdot 10^{-5}$ & $3.38 (6) \cdot 10^{-5}$ & $1.43 (5) \cdot
10^{-5}$ & $5.85 (6) \cdot 10^{-6}$ \\
15 & $2.43 (3) \cdot 10^{-5}$ & $1.09 (2) \cdot 10^{-5}$ & $4.62 (6) \cdot
10^{-6}$ & $2.19 (7) \cdot 10^{-6}$ \\
20 & $3.51 (5) \cdot 10^{-6}$ & $1.62 (4) \cdot 10^{-6}$ & $7.51 (6) \cdot
10^{-7}$ & $3.62 (7) \cdot 10^{-7}$ \\
25 & $5.37 (5) \cdot 10^{-7}$ & $2.54 (1) \cdot 10^{-7}$ & $
                     $ &                          \\
30 & $8.98 (7) \cdot 10^{-8}$ & $                      $ & $
                     $ &                          \\
\end{tabular}
\end{center}
\label{Dtable}
\end{table}

The first step in our analysis of the data is to demonstrate the
predicted exponential dependency on $N$ as given by (\ref{diff}).  In
Fig.~2 a semi-log plot of the diffusion coefficients $D$ as a function
of $N$ is presented.  We found that the data of the scaled diffusion
coefficient $D \exp(0.75 f)$ leads to a fairly good collapse of the
data for large $N$, independent of the exponent $\beta$ and the
exponential factor $\alpha$, which is demonstrated in Fig.~2.  At large
$N$ the data obey the expected exponential behavior with an effective
exponential factor of $\alpha_{eff} \approx 0.37$, very similar as in
previous simulations \cite{Needs}.  As a guide to the eye, the broken
line in Fig.~2 is $\sim \exp(-0.37 N)$.

\begin{figure}
\epsfxsize=11.5cm
\epsfbox{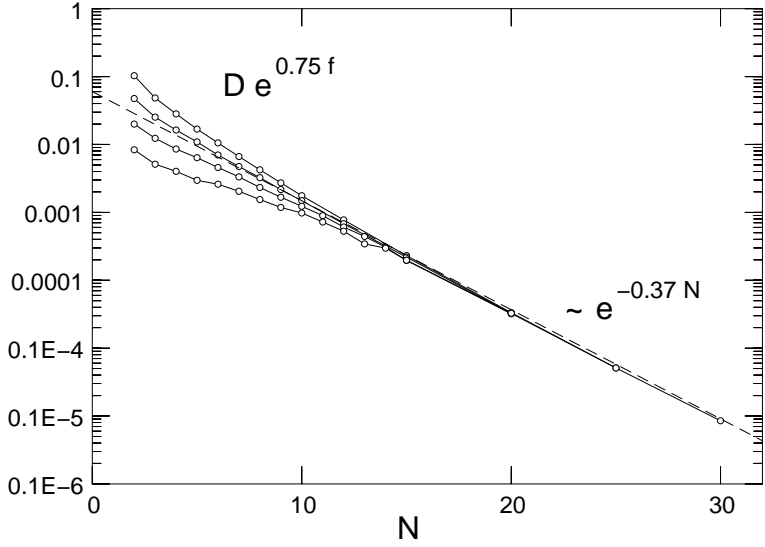}
\caption{Semi-log plot of the
Monte Carlo data of the scaled diffusion coefficients $D {\rm e}^{0.75 f}$
for $3 \leq f \leq 6$ and $2 \leq N \leq 30$.
The broken line corresponds to $\sim {\rm e}^{-0.37 N}$.}
\label{fig2}
\end{figure}

Assuming this exponential law we have analyzed the pre-exponential
power law of (\ref{diff}). This is presented in Fig.~3. In agreement
with previous simulations \cite{Needs} and theoretical considerations
\cite{Pearson} one observes for shorter chains an effective power law
of $N^{-0.6}$, given by the broken line in Fig.~3.  However, at $N>10$
there is a significant deviation from the initial behavior.  The
statistical error, as given in Table I, cannot account for this.

\begin{figure}
\epsfxsize=11.5cm
\epsfbox{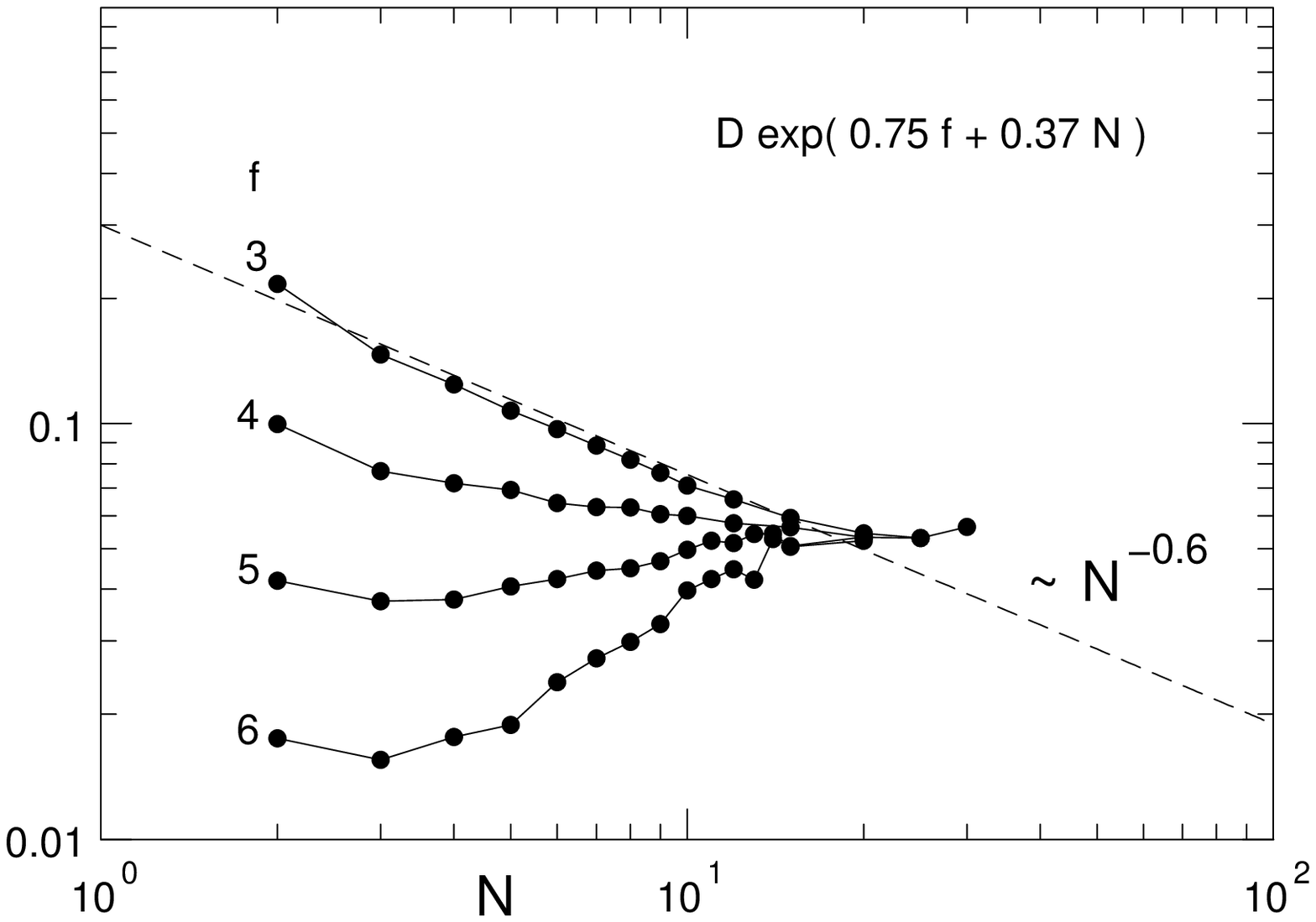}
\caption{Log-log plot of the
Monte Carlo data of the scaled diffusion coefficients
$D {\rm e}^{0.75 f + 0.37 N}$.
The broken line corresponds to $ \sim N^{-0.6}$.
The numbers at each curve denote the number of arms $f$.}
\label{fig3}
\end{figure}

Therefore, we have reexamined the data assuming the exponential factor
of Helfand and Pearson (\ref{alpha}). The corresponding analysis is
presented in Fig.~4. In this case the data indicate a pre-exponential
power law of $N^{-2}$.
Therefore, in the limit of large $N$ the data of the
diffusion coefficient seem to obey the formula
\begin{equation}
\label{d1}
D_1 = N^{-2}\;{\rm e}^{-\alpha_{HP} N -0.75f +1.7 }
\end{equation}
(the subscript $``1''$ indicates the single
star behavior trapped inside a fixed network,
in contrast to the case of concentrated
solutions of star polymers).  It is interesting to note that the
exponential term can be approximated by
$\exp\lbrack -\alpha_{HP} N -0.75(f-2) \rbrack$.

\begin{figure}
\epsfxsize=11.5cm
\epsfbox{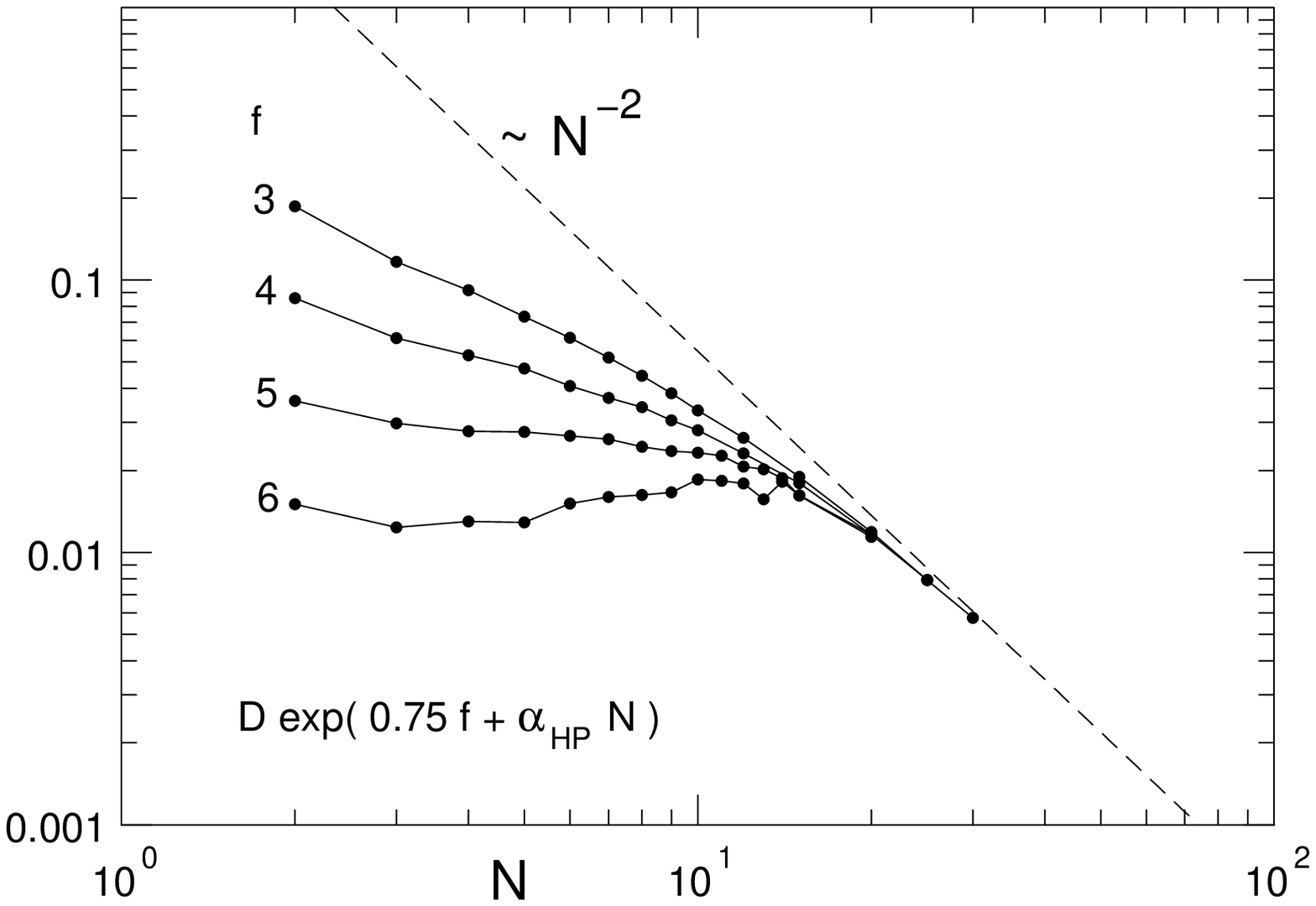}
\caption{Log-log plot of the
Monte Carlo data of the scaled diffusion coefficients
$D {\rm e}^{0.75 f + 0.29389 N}$.
The broken line corresponds to $ \sim N^{-2}$.
The numbers at each curve denote the number of arms $f$.}
\label{fig4}
\end{figure}

The normalized data $D/D_1$ are presented in Fig.~5.

\begin{figure}
\epsfxsize=11.5cm
\epsfbox{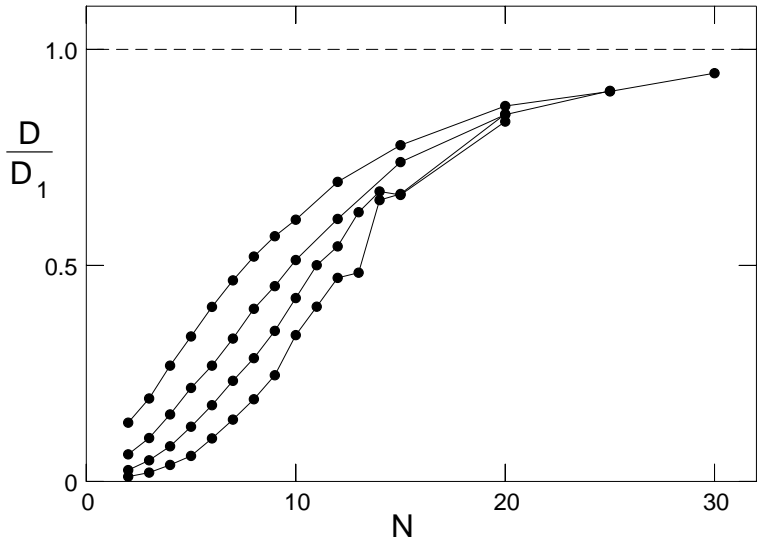}
\caption{Log-log plot of the
Monte Carlo data of the scaled diffusion coefficients
$D / D_1$, where $D_1$ is given by Eq.~(5).}
\label{fig5}
\end{figure}
\vskip 1cm

It is important to note that the $f$-dependence of $D_1$ appears as a
pre-exponential factor. Previous suggestions \cite{Graessley,Doi2}
assumed $D \sim \exp(-\alpha(f-2)N)$, based on the assumption that the
diffusion of the branch point requires simultaneous retraction of all
but two of the star arms. The latter approach cannot be reconciled
with our data. However, there is a much better agreement between our
data and recent estimates based on experiments \cite{Shull} which gave
$D \sim \exp(-0.41 f)$. These two latter findings, Monte Carlo and
experiments, strongly support earlier considerations by Rubinstein
\cite{Rubinstein86} who predicted a much weaker dependence on $f$ as
compared to the prediction $\exp(-\alpha fN)$.  Therefore the
experimental work \cite{Shull} and our analysis strongly support the
view that in order to move the center of the star, the $f-2$ arms
need not be retracted simultaneously.  Our data cannot be used to
distinguish clearly between Rubinstein's \cite{Rubinstein86} and our
formula $\exp(-0.75f)$ as given in Eq.(5).

\section{Summary and Conclusions}

We have reported on Monte Carlo results of diffusion coefficients of
lattice star polymers trapped inside a fixed network.  We found that
our data are in agreement with the Helfand-Pearson exponential factor
$\alpha = 0.29$.
However, the pre-exponential power law exponent $\beta = 2$
is different from existing theories and coincides with the
exponent for linear chains.
In contrast to existing
theoretical predictions, we found that the number of arms $f$ leads to
a pre-exponential factor of the form $\exp(-0.75 f)$.

We have not undertaken to estimate the finite size corrections to the
asymptotic formula (\ref{diff}) using our Monte Carlo data.
This would be probably quite useful comparing our results with more
refined theories proposed recently \cite{Milner}.

\vskip1cm\noindent
{\bf Acknowledgements}
We are grateful to an anonymous referee for pointing out references
\onlinecite{Shull} and \onlinecite{Rubinstein86}.



\end{document}